\documentclass[graybox]{svmult}

\usepackage{type1cm}        
\usepackage{makeidx}         
\usepackage{graphicx}        
\usepackage{multicol}        
\usepackage[bottom]{footmisc}

\usepackage{newtxtext}       %
\usepackage{newtxmath}       

\newtheorem{thm}{Theorem}

\makeindex             

\begin{document}

\title*{Microchain: a Light Hierarchical Consensus Protocol for IoT System}
\author{Ronghua Xu and Yu Chen}
\institute{Ronghua Xu \at Binghamton University, SUNY, Binghamton, NY 13902. \email{rxu22@binghamton.edu}
\and Yu Chen \at Binghamton University, SUNY, Binghamton, NY 13902. \email{ychen@binghamton.edu}}
%
%
\maketitle


\abstract{While large-scale Internet of Things (IoT) makes many new applications feasible, like Smart Cities, IoT also brings new concerns on data reliability, security and privacy. The rapid evolution in blockchain technologies, which relied on a decentralized, immutable and distributed ledger system for transaction data auditing, provides a prospective solution to address the issues in IoT. The blockchain and smart contract enabled security mechanism for IoT applications have attract an increasing interests from both academia and industry. However, integrating cryptocurrency-oriented blockchain technologies into IoT systems meets tremendous challenges on scalability, storage capacity, security and privacy. Particularly, the performance of blockchain networks significantly relies on the performance of consensus mechanisms, e.g., in terms of data confidentiality, transactions throughput, and network scalability. In this chapter, given an in-depth review of state-of-the-art blockchain networks, key matrix of designing consensus mechanism for IoT networks are identified in terms of throughput, scalability and security. To demonstrate a case study on designing scalable, lightweight blockchain protocols for IoT systems, a Microchain framework is introduced and a proof-of-concept prototype is implemented in a physical network environment. The experimental results verify the feasibility of integrating the Microchain into IoT systems.}

\section{Introduction}
\label{sec:introduct}
With the proliferation of Internet of Things (IoT), a large volume of smart devices is connected to the Internet at an unprecedented scale. The prevalence of IoT devices has changed human lives by ubiquitously providing applications and services that have revolutionized transportation, healthcare, industrial automation, emergency response, and so on. For instance, thanks to the rapid advances in IoT and edge-fog-cloud computing technologies, which are among of hot research topics in Smart Cities, Smart Public Safety (SPS) system has become feasible by integrating heterogeneous computing devices and different types of networks to collaboratively provide seamless public safety services for communities and the society \cite{nikouei2019decentralized}.

With an ever-increasing presence of IoT-based smart applications and their ubiquitous visibility from the Internet, the highly connected smart IoT devices with a huge volume of generated transaction data incur more concerns on security and privacy \cite{chen2018smart}. IoT systems are deployed in a distributed network environment that consists of a large number of devices with high heterogeneity and dynamics. The heterogeneity and resource constraint at the edge networks necessitate a scalable, flexible and lightweight system architecture that supports fast development and easy deployment with multiple application vendors using non-standard development technologies. Furthermore, those smart devices are geographically scattered across the near-site edge networks and managed by fragmented service providers that enforce different security policies. Thus, traditional security policies on a centralized authority basis, which suffer from the performance bottlenecks or single point of failures, are not efficient and suitable to address the performance and security challenges in IoT systems.

Recently, designing new decentralized security mechanisms for distributed network applications becomes one of the most intensively studied topics both in academia and industry. Blockchain, which acts as the fundamental protocol of Bitcoin \cite{nakamoto2008bitcoin}, has demonstrated great potential to revolutionize the fundamentals of information technology (IT) due to many attractive properties, such as decentralization and transparency \cite{novo2018blockchain}. Essentially, the blockchain is a public ledger based on consensus rules to provide a verifiable, append-only chained data structure of transactions. Blockchain uses a decentralized architecture that does not rely on a centralized authority, so that the data can be stored and updated distributively under a peer-to-peer network. It improves system availability and mitigates single point failure problem compared to a centralized architecture. 

In a blockchain network, a consensus mechanism is enforced on a large amount of distributed nodes called miners to maintain the sanctity of the data recorded on the blocks. The transactions are approved by miners and recorded in the time-stamped blocks, where each block is identified by a cryptographic hash and chained to preceding blocks in a chronological order. Therefore, multiple participants can access and make changes to the shared public ledger stored worldwide on distributed nodes maintained by “miner-accountants”, as opposed to establishing and maintaining trust with a transaction counter-party or a third-party intermediary. Thus, blockchain is an ideal decentralized architecture to ensure distributed transactions among all participants in a trustless environment, like edge-fog-edge computing based IoT applications under heterogeneous network environment.

Recently, there are many reported efforts that aims at addressing security issues in IoT systems leveraging the blockchain and smart contract enabled mechanisms for IoT-based applications. For example, public safety system \cite{xu2019blendmas}, smart surveillance system \cite{nagothu2018microservice,nikouei2018real}, social credit system \cite{lin2019enhance,lin2019blockchain, xu2018constructing}, decentralized data market \cite{xu2019blendsm}, space and avionics systems \cite{blasch2019blockchain, xu2018exploration}, biometric imaging data processing \cite{xu2019decentralized}, identification authentication and access control \cite{xu2018blendcac, xu2018smartcac}. All these reported researches have verified that blockchain and smart contract together are promising to provide a decentralized security mechanism to IoT systems. They have also shown that, however, directly integrating existing cryptocurrency-oriented blockchain technologies into IoT systems is hindered by several challenges in terms of scalability, computing intensity, storage capacity, data security, and privacy preservation.

The efforts in blockchain-enabled services for IoT system face critical challenges in designing blockchain network in terms of high quality of service, data confidentiality and privacy-awareness. Particularly, the performance of blockchain networks significantly relies on the efficiency of the consensus mechanisms, e.g., in terms of data consistency, speed of consensus finality, robustness to arbitrarily behaving nodes, and network scalability \cite{wang2019survey}. Unfortunately, existing blockchain protocols are mainly designed for cryptocurrency, and they are not suitable to be directly embedded into IoT scenarios.

To evaluate the challenges in designing blockchain protocols for IoT systems, this chapter provides a comprehensive overview on present blockchain networks regarding cryptographic technologies and incentive mechanisms. This chapter is organized as follows. Section \ref{sec:overview} provides an overview of blockchain fabric in the angles of system design and implementation. Section \ref{sec:consensus} explains the basics of classic fault-tolerant consensus in distributed systems, and highlights two most popular blockchain consensus protocols: BFT-based consensus and Nakamoto consensus. Given the analysis on existing issues of classic consensus protocols, Section \ref{sec:new_consessus} discusses several emerging consensus protocols that improve performance and security in blockchain networks. General challenges on integrating blockchain with IoT are identified in Section \ref{sec:challenges} based on the state-of-the-art developments. In Section \ref{sec:microchain}, Microchain, a hybrid blockchain architecture, is introduced as a case study on designing scalable, lightweight blockchain protocols for IoT systems. Section \ref{sec:conclusions} concludes this chapter and summarizes the future research opportunities for blockchain technology within the context of IoT systems.

\section{An Overview of Blockchain Fabric}
\label{sec:overview}
Compared to the traditional distributed computing paradigm with a clear client-server model, a blockhchain network allows every participant to be both a client (to issue transactions) and a server (to validate and finalize transactions) \cite{xiao2019survey}. Each participant could maintain a local view of the distributed ledger, which contains valid transactions and data. The ledger should be consistent with other nodes across the network given a underlying consensus protocol in the blockchain. The core task of a blockchain network is to ensure that the trustless nodes in the network reach agreements upon a single tamper-proof record of transactions \cite{wang2019survey}. The decentralized network architecture and the fault-tolerance enabled consensus protocol allows the blockchain to be a prospective infrastructure for distributed services and applications. 

\begin{figure}[t]
    \centering
        \includegraphics[width=0.95\textwidth]{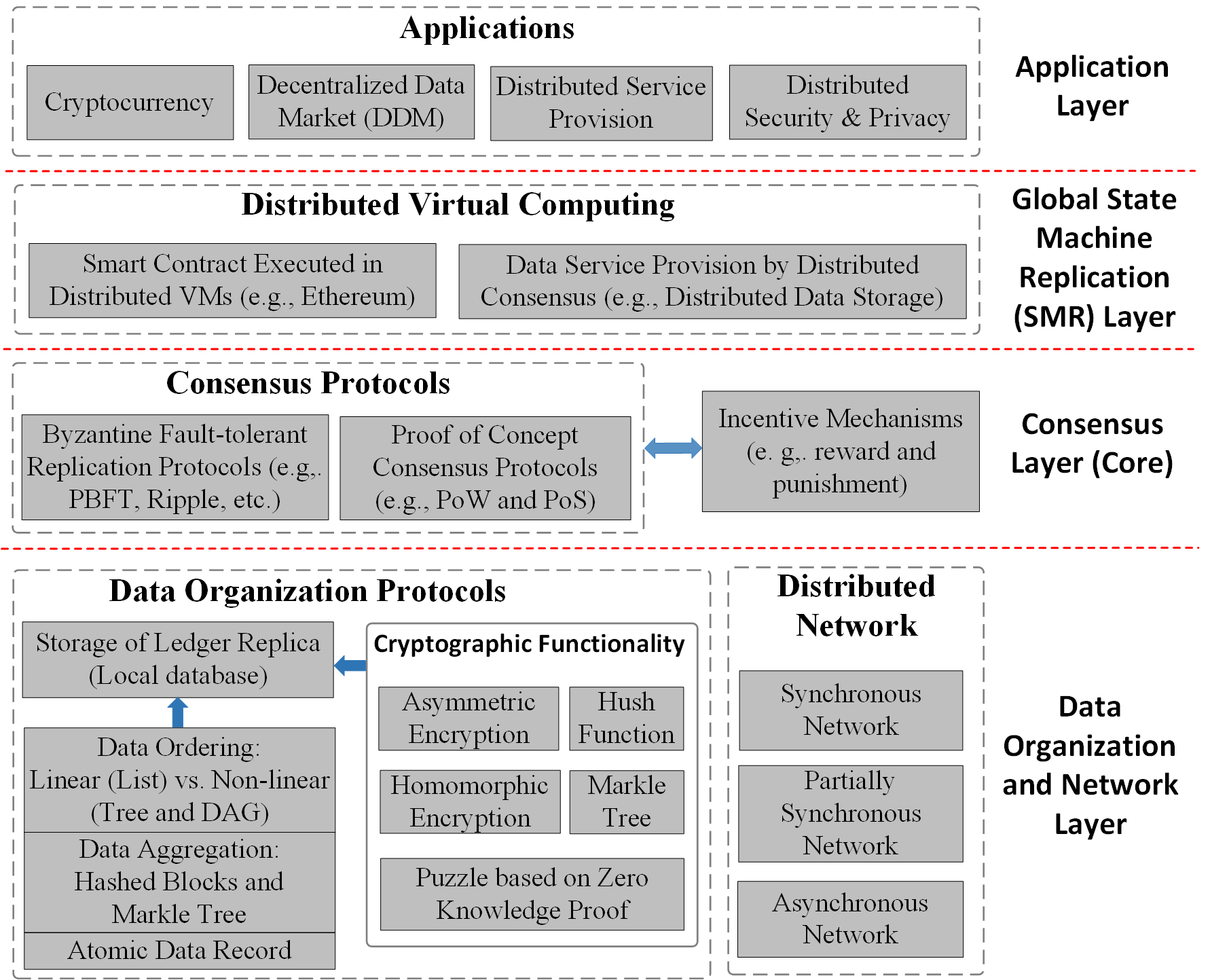}
    \caption{An overview of blockchain fabric implementation.}
    \label{fig:bc_ovewview}
    \vspace{-10pt}
\end{figure}

Figure \ref{fig:bc_ovewview} provides an overview of the blockchain infrastructure from the perspective of system level design and implementation. The application layer, which is on top of the global state machine replication (SMR) layer, exerts smart contracts to build a wide range of applications. The global SMR layer serves as a basic service level of the blockchain to support distributed computing functions for upper applications. The consensus layer acts as a core layer in the whole system by executing the consensus protocol to ensure tamper-proof of the distributed ledger and the SMR. The main function of the data organization and network layer is to identify an optimized data representation and an efficient cryptography to improve the performance of the blockchain given a certain network environment.    

\subsection{Application Layer}

In general, the emerging blockchain-enabled applications could be divided into two categories: (1) cryptocurrency and payment; and (2) the service provision based on blockchain. In short, cryptographic currencies date back to Chaum’s proposal for “untraceable payments” in 1983, a system involving bank-issued cash in the form of blindly signed coins \cite{bonneau2015sok}. Bitcoin \cite{nakamoto2008bitcoin} has been considered as the most successful cryptocurrency since it was developed in 2008. Inspired by Bitcoin's success, other cryptocurrency, like Ethereum \cite{ehtereum} and Litecoin \cite{litecoin}, has been springing out to join the competition in cryptocurrency market. Thus, cryptocurrency is a classical blockchain application, which tries to build a decentralized digital payment system based on cryptocurrency technology.

The blockchain-based service provision usually utilizes special characteristics of blockchain networks, such as self-organization, decentralization, and security, to guarantee target features provided in their respective services. A decentralized data marketplace (DDM) could use blockchain and smart contract technologies to provide secured transactions and data integrity in a self-organized big data marketplace \cite{ramachandran2018towards, xu2019blendsm}. Another prospective application is distributed security service infrastructure to provides flexible and scalable data access control and privacy preservation for applications, like public social security and smart public safety system. 

\subsection{Global State Machine Replication (SMR) Layer}
The global SMR layer acts as a distributed virtual computing service based on a smart contract or distributed data service provision. The consensus mechanism ensures all participants have agreements in the same transaction within a fault tolerance threshold. Therefore, the global SMR is achieved through the secure updating of the distributed ledger maintained by the blockchain network. Emerging from the intelligent property, a \textit{smart contract} allows users to achieve agreements among parties through a blockchain network. By using cryptographic and security mechanisms, a smart contract combines protocols with user interfaces to formalize and secure the relationships over computer networks \cite{szabo1997formalizing}. A smart contract includes a collection of pre-defined instructions and data that have been saved at a specific address of the blockchain as a Merkle hash tree, which is a constructed bottom-to-up binary tree data structure. Through exposing the public functions or application binary interfaces (ABIs), a smart contract interacts with users to offer the predefined business logic or contract agreement. Owing to the encapsulation of predefined operational logic and public exposed ABIs, smart contract is an ideal decentralized app (Dapp) backbone to support upper level applications. 

\subsection{Consensus Layer}
Given a cryptographic data organization and a blockchain network setting, the consensus layer provides the core functionality to maintain data integrity, consistence and order of data in the distributed ledger across the trustless network. From the prospective of consensus design, the consensus protocol and incentive mechanism are two key parts to ensure the performance and security of a blockchain network. The Byzantine Fault Tolerant (BFT) replication consensus protocols, like PBFT \cite{castro1999practical} and Ripple \cite{schwartz2014ripple}, execute the consensus algorithm among a small group of nodes which are authenticated by the network administrator. They are well adopted in the permissioned blockchain network in which the access control strategies for network management are enforced. On the contrary, the Proof-of-Concept (PoC) consensus protocols, like PoW and PoS, utilizes probabilistic finality schemes to achieve the consensus agreement in a open access network environment. Therefore, they provide core consensus frameworks in permissionless blockchain networks, e.g. Bitcoin and Ethereum. Other consensus protocols borrow ideas form two basic consensus mechanisms to build hybrid variants. 

The consensus protocol guarantees achieving the agreement among the nodes. However, it is expected to be \emph{Incentive Compatible} to prevent disagreements among nodes in a network because of faults or dishonest participants. Therefore, design of the consensus protocol (mining mechanism) relies on both cryptographic technologies and incentive mechanisms. The incentive compatibility of the protocol has been openly questioned from game theoretical perspectives, so that game theory are well used to analyze the strategies of consensus participants \cite{liu2019survey}. 

\subsection{Data Organization and Network Layer}

The data organization protocols use a set of cryptographic functionality to establish a unique identity and provide a fundamental protection of data confidentiality, integrity and privacy in the blockchain network. For example, asymmetric encryption could generate the public key and private key pairs. The public key is used to create the blockchain address and encrypted transactions, while the private keys could be used to create the digital signature for future verification. As the atomic data structure of a blockchain, transactions are created and signed by users or smart contracts. Each miner collects transactions on the network, and proposes blocks, where hashed transactions are represented by a Merkle tree \cite{merkle1987digital}. The blocks are organized as the "chains of block" in a chronological order. In general, the data order uses a linear linked blocks structure. To improve the processing efficiency, network scalability and security, the linear data organization framework has been expanded into the nonlinear forms, such as trees \cite{kiayias2017trees} or graphs of blocks \cite{sompolinsky2016spectre}. Each participants use its storage of ledger replica as local database to join the mining task, and regularly synchronize local ledger replica with the globalledger.

Network synchronous setting greatly influences the blockchain network in terms of consensus protocols and data representation. The synchronous network environment is a basic assumption of designing distributed consensus protocol. In a synchronous network, all messages could be delivered successfully within a fix upper bound on the time. While in a partially synchronous network, all operations and messages delivery are loosely coordinated based on an uncertain upper bound. In a asynchronous network, the upper bound on the time does not exist so that where is not a delay guarantee on message delivery.

\section{Distributed Consensus Protocols and Algorithms}
\label{sec:consensus}
In a distributed system, all participants cooperate with each others to achieve a common goal in spite of geographically separated locations. Since each node could be prone to system faults and communication channels suffer from adversarial attacks, consensus mechanism allows that the participants still can reach agreement on global state in the presence of component failures, either \emph{crash failure} or \emph{Byzantine failure}. The crash failure happens when the host system of participant abruptly stop functioning and cannot resume by itself. While Byzantine failure is caused by system malfunctions or malicious behaviors, such as sending contradictory messages to partners or withholding messages. Therefore, the consensus protocol is aimed to solve fault-tolerant problems in distributed system scenarios.  

A consensus protocol defines a set of rules for message passing and processing for all networked components to reach agreement on a common subject \cite{xiao2019distributed}. A messaging passing rule specifies the way of messages broadcasting and relaying among system components. A processing rule defines how a component changes its internal state as receiving these valid messages. The goal of consensus is reached as long as all non-faulty participants make agreement on a target subject. In general, the tolerant number of faulty nodes in a network is used to measure the strength of a consensus protocol from security's perspective. Given two failure types, the fault tolerant problems are divided into Crash-Fault Tolerant (CFT) and Byzantine-Fault Tolerant (BFT). A consensus protocol that tolerates at least one crash failure is called CFT, while BFT requires that a consensus protocol can tolerate at least one Byzantine failure. In terms of failures, crash failure is considered as a benign case while Byzantine failure is considered the worst case. Therefore, a BFT consensus is naturally CFT \cite{xiao2019distributed}. In a more precise way, BFT consensus protocol must satisfy the following properties:

\begin{itemize}
\item \emph{Validity (Correctness)}: If a honest node receives a valid common replicate proposed by other nodes, this common replicate should be accepted into the blockchain. 

\item \emph{Agreement (Consistency)}: All the honest nodes should update their local replicates of the blockchain with the block header of confirmed global blockchain.

\item \emph{Termination (Liveness)}: Every honest node should either discard or accept new transactions into the blockchain, and all transactions originated from the honest nodes will be eventually confirmed.

\item \emph{Integrity (Total Order)}: All honest nodes should accept the same chronological order of transactions which are correctly appended to the hash-chained blockchain.
\end{itemize}

In a consensus algorithm, validity, integrity and agreement define the consensus safety properties, while termination defines its liveness property. Given variant consensus protocols, the blockchain networks could be categorized into permissionless blockchain (e.g., Nakamoto Consensus Protocols), permissioned blockchain (e.g., Practical Byzantine Fault Tolerant Consensus) and hybrid blockchain. The remainder of this section will focus on underlying consensus protocols for blockchain in terms of the consensus goal and network model.

\subsection{Byzantine Fault Tolerant Consensus}
The classic consensus in a distributed system can be expressed abstractly as a Byzantine General Problem \cite{lamport1982byzantine}, which copes with single value agreement among different parts of a system given failure of communication or conflicting information. Formally, the Byzantine General Problem can be explained in a message-passing system with $N$ participants $p_i \in P$, where $i \in (1,2,...,N)$. They are geographically distributed and inter-connected by communication links. They communicate with each others only through broadcasting messages across the network to make agreement on a common plan of action: $a$ (attack) or $r$ (retreat). To achieve a agreement on a single value, each participant broadcast his/her vote for $a$ or $r$ and makes his/her decision locally based on the received votes. 

Due to the Byzantine failure, some dishonest participants $f$ attempt to prevent consensus by sending contradicting votes to different nodes. Therefore, in a network including $N$ nodes, the ultimate goal is that all loyal participants still agree on the consistent action in spite of the Byzantine failure. It requires the super-majority of the participants must be honest, which means $N-f > 2f$. So that we can get $N \geqslant 3f+1$. This can be defined as Theorem \ref{thm1}.

\begin{thm} \label{thm1}
1. In a message-passing system with $n$ nodes, if there are $f$ Byzantine nodes and $n \leqslant 3f$,  then no solution for system to achieve consensus goal. 
\end{thm}

\begin{figure}[t]
    \centering
        \includegraphics[width=0.65\textwidth]{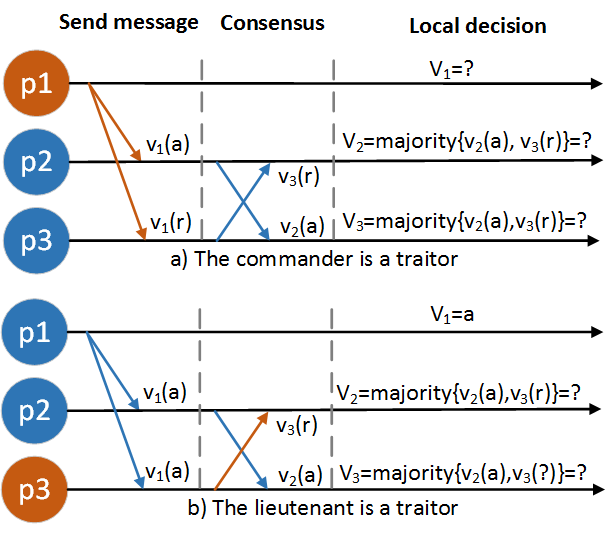}
    \caption{Example for Byzantine Problem in a three participants network.}
    \label{fig:byzantine_problem}
    \vspace{-10pt}
\end{figure}

Theorem \ref{thm1} has been conveniently proved by contradiction in Pease, M's work \cite{pease1980reaching}. Figure \ref{fig:byzantine_problem} illustrates the Theorem \ref{thm1} in a three nodes scenario with a single Byzantine node. We specify p1 as the commander, and p2, p3 are lieutenants. A traitor could be the commander or a lieutenant. In Fig. \ref{fig:byzantine_problem}a), p1 send contradictory actions a (attack) and r (retreat) to p2 and p3, respectively. Since each lieutenants will obey the order from commander, so that p2 decide to "attack" while p3 start "retreat". It violates the IC1 that all loyal lieutenants must obey the same order, then consensus goal failed. In such a scenario, both p2 and p3 cannot identify whether the commander is a traitor. Figure \ref{fig:byzantine_problem}b) demonstrate a scenario in which a lieutenant is traitor. Since p3 is a traitor and always makes a different action from p2, so that all lieutenants cannot agree on the same action. Thus, consensus goal cannot be achieved. 

\subsubsection{Oral Messaging (OM) Algorithm for BFT Consensus}

The Oral Messaging (OM) algorithm was firstly proposed as a solution to the original Byzantine generals problem \cite{lamport1982byzantine}. It assumes that there is a leader who acts as the "commander" to trigger the OM process and the other $N-1$ participants are "lieutenants" who orally pass around messages they received. Given assumptions that at most $f$ participants are Byzantine-fault nodes, $OM(f)$ could achieve BFT consensus if $f$ satisfies the Theorem \ref{thm1}, so that $N \geqslant 3f+1$. The definition of an oral message is based on the following assumptions \cite{lamport1982byzantine}:
\begin{itemize}
\item A1. Every sent message is delivered correctly.
\item A2. The receiver of a message knows the sender.
\item A3. The absence of a message can be detected.
\end{itemize}

Assumptions A1 and A2 are mainly to prevent a traitor's interference on communication between each nodes. A1 specifies that the network is synchronous so that the message can be delivered on time without error. A2 could prevent traitor from confusing others by introducing a spurious message. A3 prevents a traitor from keeping silence by simply withholding message.

\begin{figure}[t]
    \centering
        \includegraphics[width=0.75\textwidth]{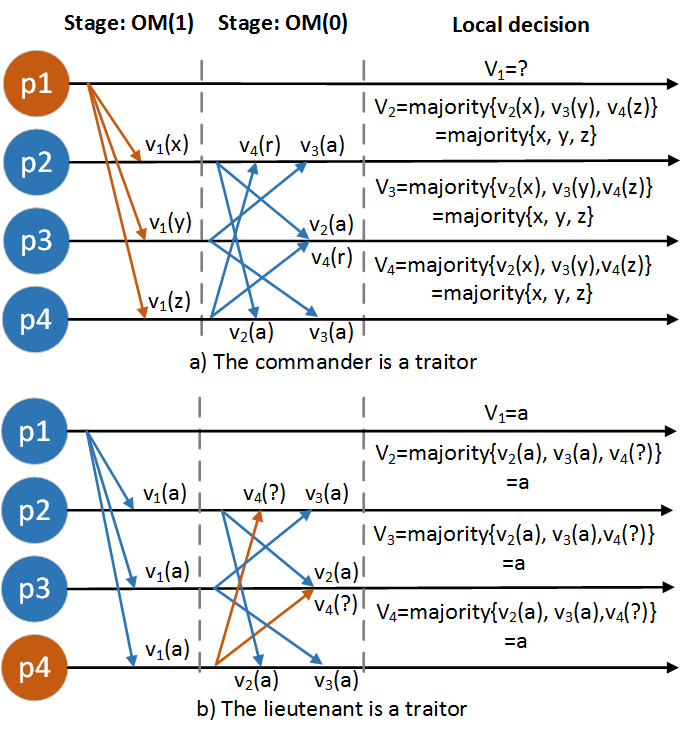}
    \caption{Example for Oral Messaging (OM) Algorithm BFT Consensus.}
    \label{fig:BFT_OM}
    \vspace{-10pt}
\end{figure}

Figure \ref{fig:BFT_OM} demonstrates how OM achieves consensus goal in a four-nodes system including only one Byzantine-fault node.  We specify p1 as commander, and p2, p3, p4 are lieutenants. A traitor could be commander or lieutenant. As a lead, p1 firstly starts to send a order message to other 3 lieutenant. In consensus stage, p2, p3 and p4 share their received message from p1. Given received messages from other lieutenants, each lieutenant makes a local decision based on the output of a function $majority$, which is defined as: $majority(v_1, v_2,...,v_{n-1})=v$, where $v$ is the majority value of $v_i$.

Figure \ref{fig:BFT_OM}b) illustrates a scenario in which a lieutenant is the traitor. In the first step of OM(1), commander p1 starts to send same order message $v_1(a)$ to other three lieutenants. In the second stage, p2, p2 and p4 use the trivial algorithm OM(0) to send their order messages to other two lieutenants. Finally, given received order messages $(v_2(a), v_3(a), v_4(?))$, each lieutenant could use $majority$ function to calculate the correct value $a=majority(a,a,?)$. Figure \ref{fig:BFT_OM}a) shows a scenario in which a commander is the traitor. Since p1 is a traitor, arbitrary value x, y and z are sent to other loyal lieutenants. Finally, each lieutenant received same order messages set $(v_2(x), v_3(y), v_4(z))$. Therefore, they all get same value $v=majority(x,y,z)$. No matter which algorithm used by $majority$, like average or maximum, they always could make agreement on a common action. 

The oral messaging algorithm $OM(f)$ is executed in a recursive fashion with $f + 1$ rounds. At the end of $OM(0)$, every lieutenant has exactly the same set of vote value calculated by $majority$. Thus, $OM(f)$ algorithm has $O(N^{f+1})$ complexity, which is impractical when $N$ is large.

\subsubsection{Practical BFT Consensus}
The classical BFT provides a solution to the single-value consensus in a synchronous network. However, the correctness of a typical distributed system not only requires every single data message is processed correctly, but it also means the processed results should satisfy the total ordering requirement. Moreover, the real-world distributed computing system relies on a partial synchronous network, or even asynchronous network. Therefore, the classical BFT consensus cannot address complex issues in a real-world distributed computing network. 

Since the sequential operations could be defined as state machines, which consists of state variables, which encode the state and commands, which transform its state \cite{schneider1990implementing}. Therefore, SMR is widely used as an active replication to ensure the data ordering consensus. The consensus protocol could be executed as a deterministic state machine, which is deployed on multiple distributed servers. Given the inputs, the state machine could change states and produce outputs in an organized manner. Through replicating the state machine across the server replicas, the SMR provides a fault-tolerant solution to the consensus protocol in distributed networks. There are two basic service requirements for SMR based protocol:\emph{liveness} and \emph{safety}. safety requires that all replicas must execute the same sequence of operations, while liveness ensures that all valid requests are executed within the consensus round. 

The next sections will introduce two SMR based fault tolerant consensus protocols for distributed computing network: Viewstamped Replication (VR) and Piratical Byzantine Fault Tolerance (PBFT).

\subsubsection{Viewstamped Replication (VR)}

The original Viewstamped Replication (VR) protocol was firstly developed in 1980s \cite{oki1988viewstamped}, and a updated was present in 2012 \cite{liskov2012viewstamped}. The VR protocol is aimed to use state machine replication to address fault tolerance issues in distributed systems. VR works in an asynchronous network like the Internet, and handles failures in which nodes fail by crashing \cite{oki1988viewstamped}. In a VR system that includes $N$ replicas, one replica works as the \emph{primary} and other $N-1$ replicas are \emph{backups}. The primary is responsible for ordering client's request while the backups simply accept orders collected by the primary. Each replica operates a local state machine with pre-defined state variables. VR ensures reliability and availability when no more than a threshold of $f$ replicas are faulty \cite{liskov2012viewstamped}. The total replicas $N$ in VR system should be no less than $2f+1$, which is the minimum replicas to ensure CFT in a asynchronous network. VR uses three sub-protocols to provide correctness:

\begin{figure}[t]
    \centering
        \includegraphics[width=0.75\textwidth]{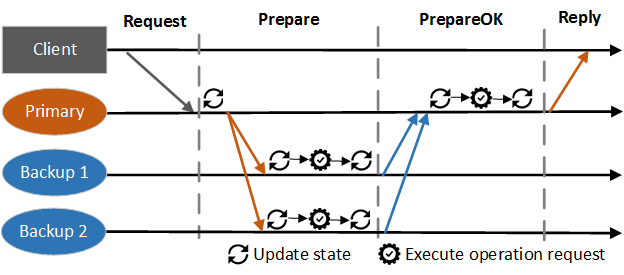}
    \caption{The normal operation sub-protocol of VR in a 3-replicas system.}
    \label{fig:VR}
    \vspace{-10pt}
\end{figure}

\begin{itemize}
\item \emph{Normal protocol}: the normal operation processes user's requests when the primary is not faulty and other backups hold the same view. Figure \ref{fig:VR} describes the normal operation workflow in a VR system with three replicas. A client sends an operation request to the primary at the beginning of a protocol run. On receiving a request message, the primary starts the \emph{Prepare} stage by firstly updating its local state, then it forwards the request to all backups using \emph{Prepare} message. Upon receiving the \emph{Prepare} message, each backup locally executes the operation request from the primary and then update its state with the given executing results. Finally, every backup sends a \emph{PrepareOK} message to the primary to notify that it has finished the operation and updated the state. After receiving $f$ \emph{PrepareOK} messages, the primary starts the \emph{PrepareOK} stage by executing the operation and updating its state accordingly. Then, the primary send a \emph{Reply} message back to the client to wrap up the replication session.

\item \emph{View change protocol}: This sub-protocol is designed to address the failure of the primary. Since backups only accept the order request from the primary, if a timeout expires without receiving the Prepare message from the primary, a backup can launch a view change request. After detecting condition for view change, a backup updates its status to VIEW-CHANGE, and sends a \emph{StartViewChange} message to other replicas. When a replica collects at least $f$ \emph{StartViewChange} messages, it sends a \emph{DoTheViewChange} message to the replica who launches the view change. When the new primary receives at least of $f+1$ \emph{DoTheViewChange} messages, it updates its state accordingly, and sends a \emph{StartView} message to notify other replicas to finish the view change process.  

\item \emph{Recovery protocol}: When a replica recovers from a crash, it cannot participant in a normal processing and view changes due to the out-of-dated state. To start the recovery protocol, a recovering replica sends a \emph{Recovery} message to all other replicas. Each replica responds with a \emph{RecoveryResponse} message indicating the current view and state. Upon received at lease $f + 1$ \emph{RecoveryResponse} messages, the recovering replica updates its local state accordingly.
\end{itemize}

Since VR is only a CFT based replication protocol which requires that total replicas should satisfy $N \geqslant 2f+1$ to prevent crash failures, it does not handle Byzantine failures as BFT consensus does. Given analysis on communication overhead on normal operation, the message complexity for VR is $O(N)$.

\subsubsection{Piratical Byzantine Fault Tolerance (PBFT)}

Since malicious attacks and software failures are increasingly common in distributed networks, both the primary and backups in the replication system are vulnerable to Byzantine failures. However, previous consensus protocols, either assumed a synchronous network with high communication complexity, like BFT-OM algorithm \cite{lamport1982byzantine}, or they cannot tolerant Byzantine failures, like VR \cite{liskov2012viewstamped, oki1988viewstamped}. To provide a practical and efficient BFT protocol under asynchronous environments, a Piratical Byzantine Fault Tolerance (PBFT) solution is proposed that advances the VR protocol to tolerant Byzantine failures in a asynchronous network \cite{castro2002practical, castro1999practical}.

The PBFT protocol can be used to implement any deterministic replicated service with a state and some operations \cite{castro1999practical}. Given a replication system which has $N$ replicas and $f$ are Byzantine faulty nodes, the PBFT algorithm guarantees \emph{safety} under condition $N \geqslant 3f+1$, which means at most $f$ replicas are Byzantine faulty. To provide \emph{liveness}, PBFT assumes a synchronous network. Therefore, clients could evenly receive replies to their operation requests only if at most $f$ replicas are Byzantine faulty and the process delay does not grow fast than upper bounded time. Similar to the VR, PBFT utilizes the sub-protocol to implement a BFT based distributed file system: normal case operation, view changes and checkpoint protocol.  

\begin{figure}[t]
    \centering
        \includegraphics[width=0.85\textwidth]{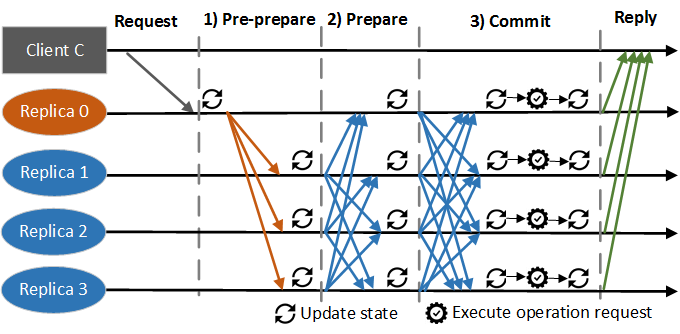}
    \caption{The normal operation protocol of PBFT in a 4-replicas system.}
    \label{fig:PBFT}
    \vspace{-10pt}
\end{figure}

\begin{figure}[t]
    \centering
        \includegraphics[width=0.85\textwidth]{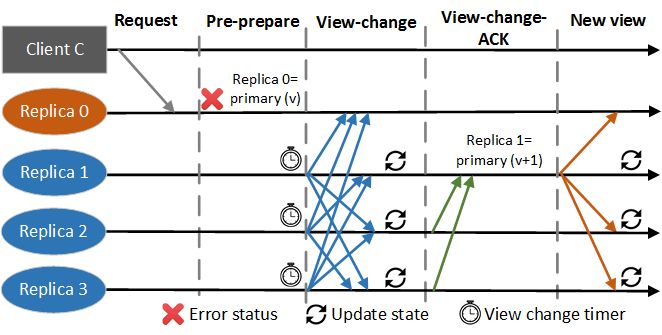}
    \caption{View changes workflow in a primary crash failure scenario.}
    \label{fig:PBFT_viewchange}
    \vspace{-10pt}
\end{figure}

\begin{itemize}
\item \emph{Normal operation}: The normal case operation uses a three phase protocol to automatically broadcast request among replicas, and it executes as session cycles with an increasing view number. Figure \ref{fig:PBFT} describes the normal case operation workflow in a PBFT system with four replicas. Replica 0 acts as a primary while other replicas are backups. A client sends operation request to the primary at the beginning of a view session to launch the three phase consensus protocol.

\textbf{1) Pre-prepare}: After receiving a request message from the Client C, the Replica 0 starts the phase one \emph{Pre-prepare} by firstly updating its local state, then it multicasts the request to other replicas using a \emph{Pre-prepare} message. Upon receiving the \emph{Pre-prepare} message, each backup checks whether the message is with valid signature and state information. If yes, the replica locally update its local status and moves to the Prepare phase by sending a \emph{Prepare} message to other replicas.

\textbf{2) Prepare}: If a backup agrees with the operation request, it votes for agreement by multicasting a \emph{Prepare} message to all replicas including the primary. After receiving at least $2f+1$ \emph{Pre-prepare} messages with the same view number and state information, a replica updates its local status accordingly and proceeds to following Commit phase by sending a \emph{Commit} message to other replicas. This phase ensures that all replicas achieve a common state before executing assigned operation requests.

\textbf{3) Commit}: Similar to phase two \emph{Prepare}, each replica firstly check the received \emph{Commit} message until there are $2f+1$ valid messages. In phase three \emph{Commit}, each replica firstly update its status to ``commit'', then locally executes the assigned operation request from the client and then update its state given executing results. This phase ensures that executed operation requests could be totally in order across cycle views. 

\textbf{Reply}: When a replica finishes the commit phase, it communicates with the client by sending a \emph{Reply} message, which encapsulates the total ordered operation request executions results and state information. The client accepts the execution results after receiving at least $2f+1$ \emph{Reply} messages.

\item \emph{View changes}: This sub-protocol ensures liveness by selecting a new primary among replicas to resume the normal operation when the current primary is in crash failure. The basic idea is that enabling a new primary could get a stable information, prepare certificates from a replica, and propagate this information to the new view \cite{castro2002practical}. Figure \ref{fig:PBFT_viewchange} illustrates the view-changes protocol from $v$ to $v+1$. Since the Replica 0 (primary\{$v$\}) is crash, the view changes are triggered by a timeout which expires without receiving valid requests from the Replica 0. After the timer triggers view changes, a replica stops receiving messages in current view and updates its status to VIEW-CHANGE. Then, it multicasts a \emph{ViewChange} message for view $v+1$ to other replicas. After receiving the \emph{ViewChange} message, a replica accepts Replica 1 as the new primary\{$v+1$\} and sends a \emph{ViewChangeACK} message to the new primary as agreement. When the new primary collects at least $2f$ \emph{ViewChangeACK} messages, it updates the state accordingly, and sends a \emph{New View} message to notify other replicas to finish the view change process and process normal operation.  

\item \emph{Checkpoint protocol}: The checkpoint protocol allows all replicas to make agreement on a stable checkpoint including the essential service state information, so that old messages from the log are safely discarded. Each replica periodically marks an executed operation request whose sequence number $h$ is divisible by a constant value (e.g., 100). Each replica broadcasts a \emph{Checkpoint} message encapsulating checkpoint information to other replicas. After receiving at least $2f+1$ \emph{Checkpoint} messages with the same $h$, it labels checkpoint $h$ as stable and records the \emph{Checkpoint} messages as the proof of correctness for stable checkpoint.

\end{itemize}

The PBFT assumes that total replicas $N \geqslant 3f+1$ when $f$ replicas are Byzantine failures. Therefore, messages from $2f+1$ non-faulty replicas are enough to achieve a super-majority consensus on voting process, like prepare and commit in normal operation. The communication overhead is evaluated in terms of three phases. Since only the primary multicasts messages to all replicas in the Pre-prepare phase, the complexity is $O(N)$. In Prepare and Commit stages, every replica broadcasts a message to other replicas, such that the complexity is $O(N^2)$.   

\subsection{Nakamoto Consensus Protocol}
To jointly address several critical issues such as pseudonymity, scalability and poor synchronization in an open-access network environment, the Nakamoto consensus protocol \cite{nakamoto2008bitcoin} is implemented as the consensus foundation of Bitcoin. The Nakamoto consensus is based on a cryptographic hash value discovery racing game called Proof-of-Work (PoW), it is widely adopted by many cryptocurrency-based blockchain networks. We use Bitcoin as the application background to explains Nakamoto Consensus and summarize its constraints and security issues. 

\subsubsection{Network Model and Permission}
Bitcoin is the most public blockchain network, and access permission is not required for a new participant interacting with other nodes in the network. In the bitcoin network, a fresh node simply connects network and searches a list of initial peers from several known DNS servers. Then it retrieves a blockchain replica from peers and uses its unique bitcoin account address to start normal operations, like sending transactions and mining blocks. 

The Bitcoin network relies on a peer-to-peer overlay network exposed to the public Internet environment. Each node executes an instance of the Nakamoto consensus protocol and maintains a replica of the blockchain. The network is modeled as an asynchronous message-passing environment without bounded message delivery delay among nodes. The FLP \cite{fischer1982impossibility} has proven the impossible result in which consensus can not be guaranteed in a fully asynchronous network with even one fault. However, Nakamoto specifies the termination requirement into \emph{Probabilistic} finality to practically circumvent the impossible consensus issue in an asynchronous network.

\subsubsection{Consensus Protocol}
The goal of Nakamoto consensus is to ensure all participants agree on a common network transaction log as a serialized blockchain. Owing to a distributed network, each node maintains a local replica and executes Nakamoto consensus protocol independently. The security of the consensus protocol requires that the majority of nodes are honest and they can correctly execute the consensus protocol. The Nakamoto protocol can be summarized into the following rules:

\begin{itemize}
\item \emph{Message Gossiping Rule}: All newly received and locally generated transactions and blocks should be multicasted to peers in a timely manner. This ensures all nodes could receive transactions and blocks in spite of the asynchronous network environment. 
\item \emph{Agreement Rule}: After receiving a block, all honest nodes should either accept or discard it based on the block's validity. In other words, all honest nodes should agree on the same blockchain if each nodes accepted the same number of blocks in its local replica.
\item \emph{Validation Rule}: All received transactions and blocks need to be validated before being appended to the blockchain or broadcasted to peers. Only valid transactions could be saved into new blocks and valid blocks encapsulating valid transactions could be accepted by the blockchain network.
\item \emph{Proof-of-Work (PoW)}: Every node has to solve a computing-intensive, time-consuming hash puzzle as a Proof-of-Work for block generation. In brief, PoW solution requires exhaustively querying a cryptographic hash function for a partial preimage generated from a candidate block \cite{wang2019survey}. The hashcode of a candidate block is expected to satisfy a pre-defined difficulty condition parameter $h$, like fixed length of bits are 0s. The PoW puzzle can be formally defined as the following equation:
\begin{equation}
\label{pow_equitation}
hash\_block = \mathcal{H}(block\_data|nonce) \leqslant D(h)
\end{equation}
where for some fixed length of bits $L$ and $D(h)=2^{L-h}$, $\mathcal{H}(\cdot)$ is a pre-defined collision-resistant cryptographic hash function that outputs hash string $L \in \{0,1\}^{\lambda}$. 
\item \emph{Longest Chain Rule}: All honest nodes should always extend proposed blocks on the longest chain that they has ever found. The longest chain rule ensures the consensus in an asynchronous network, such that all honest miners are working on a common main chain. This rule ensures probabilistic finality given a certain length of sequential blocks. 
\end{itemize}

Above rules provide safety for achieving Nakamoto consensus in a distributed network. However, incentive mechanism is also indispensable to a public blockchain network, especially for those who act as financial infrastructure, like cryptocurrency and digital payment systems. Nakamoto consensus uses block rewards and transaction fees as an incentive mechanism to encourage participants to invest computation power to join the network and make contributions. 

\subsubsection{Chain Finality and Complexity Analysis}

The PoW process defined by Eq. (\ref{pow_equitation}) is essentially a verifiable process of a weighted random coin-tossing, where the probability of winning is no longer uniformly associated with the nodes' identities but in proportion to the resources, e.g., hash rate casted by the nodes \cite{wang2019survey}. In the PoW-like leader election process, the probability of a node for winning the block generation follows:

\begin{equation}
\label{pow_probability}
p_{win\_block}(i) = \frac{w_i}{\sum_{i \in N}w_i}
\end{equation}

where $w_i$ is the shared verifiable resource node $i$ can has, such as computational power, memory and storage, etc. 

According to the longest-chain rule, blocks appended to a chain branch that is not suffix of the longest chain shall be discarded or "orphaned". As defined by Eq. (\ref{pow_probability}), if attackers have more than 50\%  of the whole network's gross hash computing power, they will have higher hash generating rate so that producing blocks faster than rest of the participants in the network. The probability of an attacker to win the longest chain by continuously generating $m$ blocks is:

\begin{equation}
\label{chain_probability}
P_{win\_chain} = (\frac{p_{win\_block}}{1-p_{win\_block}})^m
\end{equation}

The $P_{win\_chain}$ drops exponentially as $m$ increases if $p_{win\_block} < 0.5$. Therefore, if more than half of the miners are honest, it is computationally impossible for attackers to revoke a block from the blockchain. Bitcoin network specifies $m=6$ as the longest chain confirmation. Since the Nakamoto consensus protocol uses a gospel style message delivery without using all-to-all message phase like BFT, it also produces smaller communication complexity $O(N)$. 

\subsubsection{Constraints and Vulnerabilities}
The Nakamoto consensus protocol demonstrates good scalability in a trustless, open-access network environment. However, PoW also incurs several performance issues, such as limited throughput, high demand of computation and storage resources as well as unsustainable energy consumption. Furthermore, verifiable random block generation and probabilistic finality make Nakamoto consensus protocol vulnerable to several security problems, like majority attack and selfish mining.

\begin{itemize}
\item \emph{Trade-off between performance and security}: The difficulty level in the PoW process is mainly to secure the block generation process. The average block confirmation time in Bitcoin is about ten minutes, which means it can process up to seven transactions per second. It becomes throughput bottleneck when blockchain technology is applied to digital cash and payment systems, which require higher transaction throughput. Increasing block size to commit more transactions and reducing blockchain confirmation time are both solutions to improve transaction throughput capacity. However, they increase block propagation delays and lead to insufficient dissemination. Insufficient dissemination could cause higher fork rate and be vulnerable to 51\% attack and self-mining.

\item \emph{Computation intensive and energy inefficiency}: In Bitcoin network, all participants invest in more resource to win the computation-intensive puzzle-solving race. This hash rate competition not only wastes computational power to calculate a meaningless ``nonce'', but also introduce huge energy cost.

\item \emph{Majority (51\%) attack}: In Nakamoto Consensus, if attackers have more than 51\% network gross computing power, they can calculate the ``nonce'' value quicker than other nodes in the network. As a result, attackers have a higher probability to generate new blocks than others. Finally, the longer chain controlled by attackers is adopted as the main chain, while the blocks mined by honest nodes are discarded. Through 51\% attack, attackers can take control of the whole blockchain network, launch malicious activities, like double spending, revoking transactions and blocks, etc.

\item \emph{Selfish mining}: The Nakamoto consensus protocol provides 50\% fault tolerance given an assumption that all miners broadcast new blocks immediately upon successful generation. However, malicious miners or mining pool may delay broadcasting newly mined blocks but withholding the blocks, and releasing blocks at a specific time to increase their benefit. As a results, honest miners waste their computational power in verifying the already mined blocks, but malicious miners can therefore increase their probability of mining the new blocks to gain rewards. In the worst case, attackers can use selfish mining strategy to control the whole blockchain network.  
\end{itemize}

\section{Emerging Consensus Protocols}
\label{sec:new_consessus}
In general, the PoW-based Nakamoto blockchain provides scalability and probabilistic finality in partial synchronous or asynchronous networks at the cost of low throughput and high energy assumption. Whereas BFT-based blockchains offer an excellent performance and deterministic finality in permissioned networks consisting of a small numbers of authorized replicas. However, due to the synchronous environment requirement and communication complexity $O(N^2)$, traditional BFT-based consensus protocols demonstrate limited scalablility when being applied to real world applications. Interested readers are referred to \cite{vukolic2015quest, croman2016scaling} for a comprehensive evaluation. 

To address the trade-offs in terms of performance, scalability and security, new consensus protocols and blockchain schemes have been exploited both in academic community and industry. This section introduces several emerging consensus protocols that enable efficient and secured blockchain networks.

\subsection{Proof-of-Stake (PoS)}

The PoW is essentially a random leader-election process through a brute-force manner to solve a cryptographic hash puzzle problem. To improve the computational-intensive hash value calculating, Proof-of-Stake (Pos) was first proposed in Peercoin \cite{king2012ppcoin} as an alternative to PoW in blockchain community. Within a Peercoin network, a matric ``coin age'' is proposed to measure a miner’s stake by multiplying the held tokens and the holding time. The PoS kernel protocol allows a miner to use its stake to solve puzzle solution, then the probability of proposing a new block follows the stake distribution. In a PoS consensus network, each validator uses its stake, which is a amount of deposit currency, to participate the mining process. Mining block is designed as a random variable function:
\begin{equation}
\label{pos_equitation}
\mathcal{TB}(\mathcal{H}(block\_data|pk_i|s_i), \delta) \leqslant D(\delta, p_i)
\end{equation}
where for adjustable difficulty parameter $\delta$, $D(\delta, p_i)=(2^\delta-1)p_i$. $\mathcal{H}(\cdot)$ is a pre-defined collision-resistant cryptographic hash function that outputs a hash string. The $\mathcal{TB}(\cdot)$ function outputs lower bits of the hashcode. $p_i$ is the stake ratio of node $i$ to gross stake in the network, which can be represented as $p_i=\frac{s_i}{\sum_{i=1}^Ns_i}$. $N$ is the total number of nodes in the network. The bigger stakeholder has a higher chance to mine a new block.

The PoS also relies on the longest chain rule to achieve probabilistic finality in the consensus process. Compared to PoW protocols that use the brute-force hashing power to solve computation-intensive puzzle solution, PoS leverages the distribution of token ownership to simulate a verifiable random function to propose new blocks. Since the block miners only consume limited resources, PoS is also known as a process of ``virtual mining'' \cite{bonneau2015sok}. Similar to the PoW in Nakamoto consensus, as long as dishonest nodes own less than half of the total stake value in network, the probability of a block being revoked from the blockchain drops exponentially as the chain grows. In a PoS network, rational attackers have weak motivation to perform 51\% attack, since deteriorate stake value is economically devastating the benefit of attackers, who are also larger stakeholder.

\subsection{Other Proof-of-Concept (PoX) Consensus Protocols}
Under the successful framework of Nakamoto consensus protocol, a number of alternative Proof of Concept (PoX) schemes have been proposed to address existing issues of PoW in terms of security, fairness and sustainability. From a network-level perspective, PoX generally relies on a pseudo-random oracle to provide the property of verifiable unpredictability or lead-election process \cite{wang2019survey}. To address the issue of centralized computation power pool, Proof of Memory (PoM), a memory-hard PoW is adopted by ZCash \cite{hopwood2016zcash} and Ethereum \cite{buterin2014next} networks.  With the purpose of useful resource provision, the idea of ``Proof of Useful Resources'' (PoUS) has been proposed to tackle the resource wasting problem of PoW \cite{wang2019survey}. The Proof of Exercise (PoE) is proposed to replace the computation intensive searching problem in PoW with the useful ``exercise'' of matrix product problems \cite{shoker2017sustainable}. Based on the Trusted Execution Environment (TEE), Resource-Efficient Mining (REM) \cite{zhang2017rem} verifies and measures the software running in an Intel Software Guard Extensions (SGX)-protected enclave that randomly determines whether or not the work leads to a valid block proof. 

\subsection{Hybrid Consensus Protocols}
\subsubsection{Bitcoin-NG}
To improve the limited performance of permissionless consensus without undermining the unique features such as scalability, combining a scalable permissionless consensus (e.g. PoW) with a high throughput permissioned consensus (e.g. BFT) becomes a prospective approach. The Bitcoin-NG \cite{eyal2016bitcoin} is proposed to improve the performance of PoW-based Nakamoto protocol by decoupling Bitcoin’s blockchain operation into two planes: leader election and transaction serialization. The protocol divides time into sequential epochs, and only a single leader is in charge of serializing state machine transitions in each epoch. To bootstrap the transaction throughput, the protocol introduces two types of blocks, namely, the key blocks that require a PoW puzzle solution for leader election and the microblocks that require no puzzle solution and are used for transaction serialization \cite{wang2019survey}. Although Bitcoin-NG may also experience key block forks, it scales optimally with bandwidth limited only by the capacity of the individual nodes and latency limited only by the propagation time of the network \cite{eyal2016bitcoin}.

\subsubsection{BFT-style PoS}
Since the BFT consensus protocol uses a deterministic finality mechanism such that all confirmed blocks will never be tampered with or revoked from blockchain. Therefore, adopting BFT style chain finality to PoS consensus could ensure data consistency and immediately finality. BFT-style PoS has been implemented in Tendermint \cite{kwon2014tendermint}, Algorand \cite{gilad2017algorand} and Ethereum's Casper \cite{buterin2017casper}. Instead of following Nakomoto's contention-based blockchain generation process, BFT-style PoS embraces a more radical design in which the set of validators periodically finalize blocks in the main chain through BFT consensus \cite{xiao2019distributed}. Compared with tradition chain-based PoS consensus protocols, BFT-style PoS provides a deterministic finality to guarantee that finalized blocks cannot be revoked. Furthermore, deterministic finality allows for designing reward and punishment strategy to discourage malicious validators to launch attacks, such as double betting or nothing-at-stake attacks. 

\subsubsection{Parallelism in BFT Consensus}
To design a computationally-scalable Byzantine consensus protocol for blockchain, SCP \cite{luu2015scp} is proposed through incorporating BFT and sharding into blockchain consensus. The key ideas are inspired by concept of ``sharding'' \cite{croman2016scaling} in infrastructure of distributed database and cloud. Through securely establishing identities for network participants, whole network are randomly divided into several sub-committees. Each sub-committee performs a classical BFT consensus protocol to process a separate set of transactions and propose blocks in parallel. A final committee is designated to combine the outputs of sub-committees into an ordered blockchain data structure. 

To extend existing consensus protocol based on SCP, Elastico \cite{luu2016secure} is proposed to build a secure sharding protocol for open blockchains. In a epoch, the candidates of committees attempt to find a PoW puzzle solution provided a seed called ``epochRandomness'', which is a public random number string generated in the previous epoch. Elastico exhibits almost linear scalability throughput with computation capacity with roughly $O(n)$ message complexity. However, the participants have to download full blockchain data to perform the consensus task, which brings latency in bootstrapping process and storage overload on client nodes.

To enable the parallelization of both network consensus and data storage, a ``full sharding'' protocol called ``OmniLedger'' \cite{kokoris2018omniledger} is designed to provide ``statistically representative'' shards for permissionless transaction processing. OmniLedger uses a bias-resistant protocol called RandHound \cite{syta2017scalable} to generate epoch global randomness strings for sharding committees formation. To optimize trade-off between the number of shards, throughput and latency, the intra-shard consensus follows an ``Optional Trust-but-Verify Validation'' model, where optimistic validators make a provisional but unlikely-to-change commitment and core validators subsequently verify again the transactions to provide finality and ensure verifiability \cite{kokoris2018omniledger}. To secure corss-shard transactions, OmniLedger introduces a novel Byzantine Shard Atomic Commit protocol to handle atomically transactions processing across shards. Furthermore, a gradually in-and-out committee members swap strategy could reduce extra message overhead and bootstrapping the latency in shard reconstruction. Another epoch-based, two-level-BFT protocol called RapidChain \cite{zamani2018rapidchain} is proposed for scaling blockchain via full sharding. RapidChain employs block pipelining strategy to achieve very high throughputs in the intra-committee consensus. Further more, a novel gossiping protocol for large blocks reduces the large overhead on committee-to-committee communication, and ensures an efficient cross-shard transaction verification.

\section{Challenges on Integration Blockchain with IoTs}
\label{sec:challenges}
Thanks to the distributed ledger, Blockchain can enrich the IoT by providing a trusted sharing service, where information is reliable and can be traceable. Blockchain and smart contract technologies are identified as the key to solve scalability, privacy, and reliability problems related to the IoT paradigm. However, directly incorporating blockchain into the IoT is infeasible. Since the traditional blockchain networks, like bitcoin, were designed for an Internet-based scenario, where rich-resource devices like computers are participants, and the network environment is stable. Consequently, it cannot meet the requirements of IoT reality, such as constrained computation and storage, communication efficiency and energy consumption, etc. Although several recent efforts that focus on either improving the performance of PoW blockchain like Bitcoin-NG \cite{eyal2016bitcoin} or scaling classical BFT  protocols through parallelization like sharding \cite{luu2016secure}, those cryptocurrency-based blockchain solutions bring up other issues when introducing into IoT systems. The identified challenges are presented as follows:

\begin{itemize}
\item[1)] \emph{The trade-off between scalability and efficiency}: The IoT applications, such as smart surveillance system, involve a large volume of generated transaction data among users and service providers, the efficient throughput and lower latency become key metrics of designing blockchain protocol for IoT. Utilizing BFT and state-machine replication protocols can potentially improve the efficiency with the trade-off of poor scalability, which causes system security issues like being vulnerable to Sybil attack. Furthermore, IoT systems generally rely on identity registration and authentication process to enroll known participants due to legal and compliance reasons, implementing permissioned blockchain for IoT could ensure a certain level of security with the supplementary node identity management.

\item[2)] \emph{The cost for transaction confirmation and storage}: Since IoT devices are resource-constrained with limited computation and storage capacity, the high complexity consensus based on computing intensive cryptographic algorithms like PoW is not affordable to IoT applications. In addition, storing the whole blockchain history to validate the current state is not only overwhelming for storage constrained IoT devices, but also introduces longer bootstrap time when new nodes join the network. Furthermore, blockchain runs on a peer-to-peer network and consensus protocol requires frequent data transmissions and exchanges to ensure consistent records in distributed ledger. It will bring significant communication overhead on light IoT networks and extra energy consumption by data transmission. Thus, lightweight considerations, such as efficient transaction processing, optimized chain-data organization, and energy saving, etc., are critical to design new blockchain consensus algorithms for IoT systems.

\item[3)] \emph{The conflicts between transparency and privacy}: As an important characteristic of blockchain, transparency allows all participants to access blockchain data and audit the transactions. However, it brings concerns on privacy issues for some IoT systems, such as e-health and smart home, where the collected sensitive user data should be confidential and are only accessible to authorized entities. Enforcing access control mechanism to some extent encounters the transparency principle of blockchain. But for some IoT applications like supply chain management, data traceability is mandatory at the cost of transaction transparency. Thus, trade-off between transparency and privacy becomes an import factor in the design of blockchain based IoT systems.  

\item[4)] \emph{Security on IoT data and blockchain}: IoT devices are vulnerable to network attacks compared with computers and cloud service, Corrupted IoT data from compromised devices make the cast that data itself is not correct before sending transactions to blockchain network. Hence, the data finalized in blockchain is polluted. On the other hand, the blockchain consensus protocol can tolerant some certain level of Byzantine failure given Byzantine nodes are below a threshold. However, more compromised IoT devices also make the consensus vulnerable to Byzantine failure, so that data in the chain are not immutable. Security should be considered in terms of IoT data and blockchain network.
\end{itemize}.

Considering above challenges on incorporating blockchain technology into IoT systems, designing optimized blockchain fabrics empowered with light and efficient consensus protocols become a prospective solution. The following section will introduce \emph{Microchain} to demonstrate how to implement a partially decentralized, scalable and lightweight distributed ledger protocol for IoT applications. 

\section{Microchain Fabrics for IoT}
\label{sec:microchain}
To address challenges in integrating blockchain technologies into IoT systems, Microchain \cite{xu2019microchain} was proposed by designing an efficient consensus mechanism running on a small number of validators. The rationale for microchain is described as follows:

\begin{itemize}
\item \emph{Permissioned committee network}: Following the idea of delegation, microchain chooses a small subset of the nodes in the network as validators, and those validators form a final committee, called dynasty, to perform a consensus function. Permissioned networks could provide basic security primitives, such as public key infrastructure (PKI), identity authentication and access control, etc. The committee with limited validators also improves the performance of consensus protocol by reducing messages propagation delay and blocks confirmation time.  
\item \emph{Random committee election}: A random committee election ensures that committee members selection process is unpredictable. Even if attackers could compromise the current committee, the probability that adversary controls consecutive committee becomes exponentially decreased owing to the unbiased random protocol used in committee election process.     
\item \emph{Computational efficient virtual mining}: Instead of using a brute-force manner to solve a computation-intensive puzzle problem as PoW does, microchain uses an computationally efficient virtual mining way called Proof-of-Credit (PoC) to enable probabilistic block generation among committee members. For every epoch of block generation, a validator is only allowed to calculate hash value once to solve puzzle problem, and the probability that a validator could mine a block is associated with
its credit distribution in the current dynasty. Such a virtual mining process is affordable to be executed by resource-constrained IoT devices with a limited computation overhead resulted from the hash calculation and signature verification.   
\item \emph{Deterministic chain finality}: PoC offers a probabilistic block generation, however, it will inevitably produces forks during chain extension. Inspired by the idea from BFT style consensus protocols, microchain introduces a voting-based chain finality to provide deterministic finality for chained blocks. The voting-based chain finality method ensures that all finalized blocks will never be tampered with or revoked from the blockchain. 
\item \emph{Incentives compatibility}: The consensus protocol should be incentive compatible, which means that any consensus node will suffer from finical loss whenever it deviates from truthfully following the protocol. The microchain uses an incentive mechanism which is based on proper rewarding and punishment polices to offer incentive compatibility. The rewarding strategies motivate more nodes join the blockchain network to maximize their benefit, while punishment strategies are aimed to discourage participants from misbehaving or launching attacks. 
\end{itemize}

\subsection{System Architecture Design}

\begin{figure}[t]
    \centering
        \includegraphics[width=0.75\textwidth]{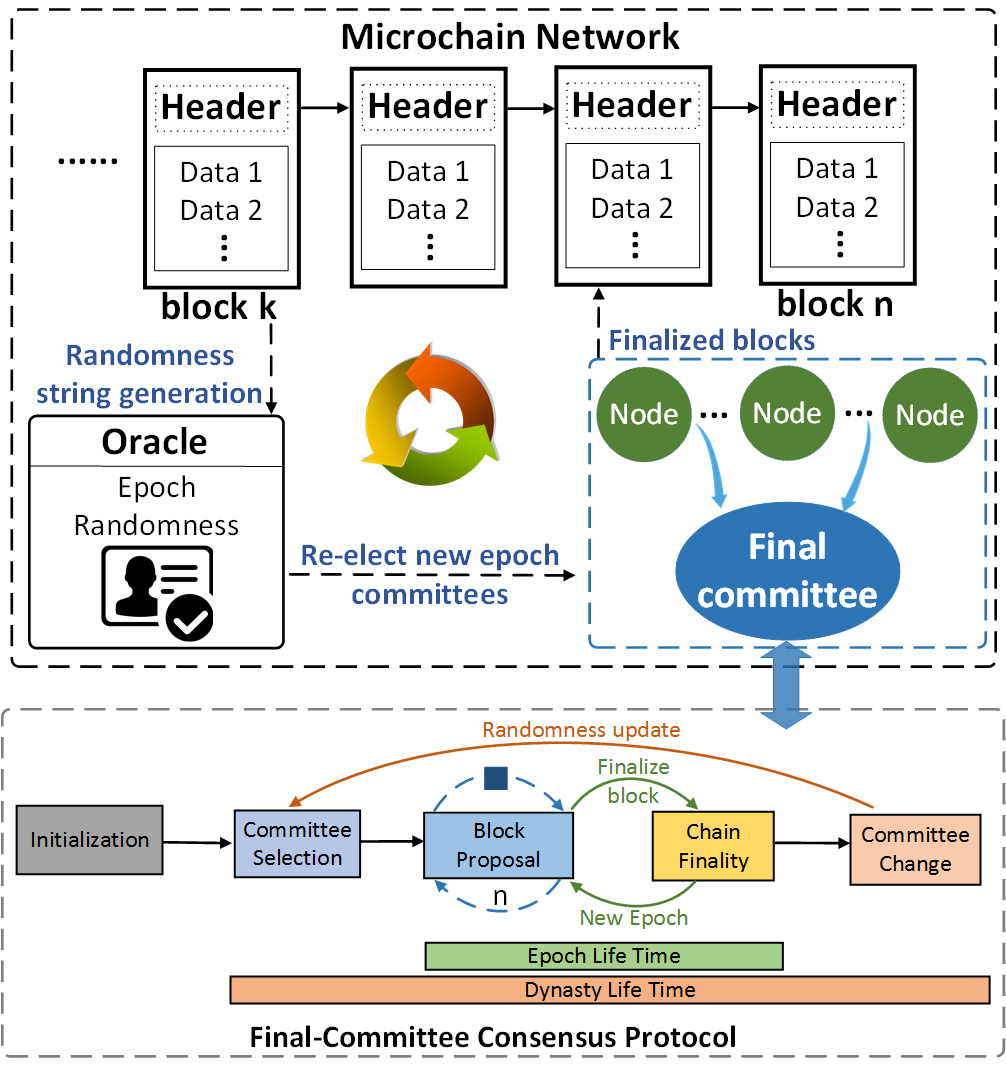}
    \caption{Microchain system overview.}
    \label{fig:microchain}
    \vspace{-10pt}
\end{figure}

The microchain network is shown in the upper part of Fig. \ref{fig:microchain}. The microchain is built on top of a permissioned network where only registered entities are authorized access to the network, allowed to interact with other validators, and contribute to consensus protocol, such as transactions propagation, block verification and mining. For dynasty epoch during a fixed time period, a final committee is responsible for key functions of consensus protocol, like transactions processing, blocks generation and chain finality. A random committee formation protocol ensures that the committee election process is unpredictable. During the lifetime of each dynasty, a final-committee consensus mechanism is responsible for proposing blocks and finalizing the chain history given an unbounded time delay \cite{xu2019microchain}. Given the assumption that a synchronous network in which operations of processes are coordinated in rounds with bounded delay constraints, microchain ensures $persistence$ and $liveness$, which are two formal proprieties of a robust distributed ledger.  

The performance and security proprieties of microchain system are relied on a light and efficient consensus protocol design. The final committee consensus protocol is illustrated in lower Fig.\ref{fig:microchain}, and key components and workflows are described as follows:

\begin{itemize}
\item \emph{Initialization}: In the initialization process, a special dynasty, which includes a group of validators specified by the administrator, acts as initial committee $D_{init}$ to initialize blockchain. Each validator creates a genesis block $B_0$ and sets the local blockchain $\mathcal{C}=B_0$ and $head=B_0$. The initial committee will work as the first dynasty of the system until the election of the next dynasty. 

\item \emph{Committee Selection}: At the beginning of the lifetime of each dynasty, the final-committee formation protocol exploits a Verifiable Random Function (VRF) based cryptographic sortition scheme \cite{gilad2017algorand} to randomly choose a subset of validators $V$ as the final-committee according to their credit weight. The selected committee members $D$ will be added to the current block, which is marked as the beginning block of the new dynasty epoch. The lifetime of dynasty epoch starts from committee selection and ends after dynasty change.

\item \emph{Block Proposal}: The block proposal mechanism uses a pure Proof-of-Stake (PoS) protocol, called Proof-of-Credit (PoC), to generate new blocks in each block proposal run. Only validators in the current dynasty can propose a new block. The probability that a validator $v_j$ could propose a block is associated with its credit distribution of the current dynasty ($pk_j, c_j$), where $pk_j$ is the public key. If validator $v_j$ could solve a puzzle in slot $sl_{t+1}$ by computing $\mathcal{H}(B_{i}, pk_j, c_j) \leq d_{cond}$ ($d_{cond}$ is difficulty condition target value), it generates new block $B_{i+1}$ and broadcasts it with a valid signature to all committee members of the current dynasty. Each committee member accepts all valid blocks in the current slot, and the verified block will be added to the local chain $\mathcal{C}$ with setting $head=B_{i+1}$.

\item \emph{Chain Finality}: At the end of an epoch, the $head$ with epoch height becomes a checkpoint that is used to resolve forks and finalize chain history. The chain finality uses a voting-based algorithm to commit checkpoint block and finalizes those already committed blocks on the main chain. The chain finality ensures that only one path, including finalized blocks, becomes the main chain, as Fig. \ref{fig:microchain} shows. Therefore, the following blocks in the new epoch are only extended on such a unique main chain. The chain fork problem is prevented by resolving conflicting checkpoints and finalizing the history of the blockchain. 

\item \emph{Committee Change}: At the end of the lifetime of a dynasty, the current committee members agree on a new dynasty randomness string. The epoch randomness string generation uses the RandShare mechanism to make an agreement on proposing the next epoch randomness string among members of the final committee. RandShare is a randomness protocol which is based on Publicly Verifiable Secret Sharing (PVSS) \cite{stadler1996publicly, schoenmakers1999simple} to ensure unbiasability, unpredictability, and availability in public randomness sharing. The proposed unbiasable and unpredictable public randomness string will be used for the committee selection process of the next dynasty lifetime.
\end{itemize}

There are two core functions in epoch lifetime of chain extension: (1) the Proof-of-Credit (PoC) protocol, which is a pure PoS mechanism, determines whether a participant is selected to propose a block given fair initial distribution of the credit stake to the committee members in a given epoch; and (2) a Voting based chain finality (VCF) mechanism could protect against fork by resolving conflicting checkpoints and finalize the history of chain data.

\subsection{Prototype Implementation and Evaluation}

To verify the proposed solution, a concept-proof prototype of microchain is implemented in Python, consisting of approximately 3000 lines of code. Flask \cite{flask} is used, which provide a micro-framework for Python application, to develop networking and web service functions. All security functions are produced by using the standard python lib cryptography \cite{pyca}. The key generation and signature are implemented over RSA, and the hash function uses SHA-256. SQLite\cite{sqlite}, which is a lightweight and embedded SQL database engine, is used to manage data such as node, block, and vote information.

The prototype is deployed on a physical network environment, including multiple nodes. Table \ref{tab:testbed} describes devices used for the experimental setup. Five validators are deployed on a desktop while other validators run on sixteen distributed Raspberry Pis to emulate an IoT environment. Each validator is only deployed on one host machine.

\begin{table}[ht]
\caption{Configuration of Experimental Nodes.} 
\label{tab:testbed}
\begin{center}       
\begin{tabular}{|l|p{3.9cm}|p{5.9cm}|} 
\hline
\rule[-1ex]{0pt}{3.5ex} \textbf{Device} & Dell Optiplex 760 & Raspberry Pi 3 Model B+ \\
\hline
\rule[-1ex]{0pt}{3.5ex} \textbf{CPU} & 3 GHz Intel Core TM (2 cores) & Broadcom ARM Cortex A53 (ARMv8), 1.4GHz \\
\hline
\rule[-1ex]{0pt}{3.5ex} \textbf{Memory} & 4GB DDR3 & 1GB SDRAM \\
\hline
\rule[-1ex]{0pt}{3.5ex} \textbf{Storage} & 250G HHD & 32GB (microSD card) \\
\hline
\rule[-1ex]{0pt}{3.5ex} \textbf{OS} & Ubuntu 16.04 & Raspbian GNU/Linux (Jessie) \\
\hline
\end{tabular}
\end{center}
\end{table}

\subsection{Network Latency}

To evaluate the network latency incurred by executing microchain on IoT devices in terms of the number of nodes, validators are deployed on 16 Raspberry Pis performing an entire round of final-committee consensus. The block size used in the test is 128KB to reduce the  influence of block sizes on network performance. The latency of committing a transaction $\mathcal{T}_{ct}$ is used for evaluating the time for all nodes of the dynasty to accept a broadcasted transaction. Since the communication complexity of broadcasting transactions is $\mathcal{O}(K)$. The latency of committing transactions is a linear scale to committee size $K$, and it varies from 162 ms to 246 ms. The latency of block proposal $\mathcal{T}_{bp}$ calculates how long the proposed blocks could be arrived and verified among validators. Since the block proposal algorithm is proportion to credit distribution $\mathcal{D}$ with expectation $E(\mathcal{D})$, the latency of block proposal is scale to communication complexity $\mathcal{O}(\frac{K^2}{E(\mathcal{D})})$, which varies from 0.5 s to 1.7 s. Finally, the latency of chain finality $\mathcal{T}_{cf}$ is the time it takes the voting process for finalizing the checkpoint block to complete among all nodes. Owning to the communication complexity $\mathcal{O}(K^2)$ in the voting-based chain finality process, The $\mathcal{T}_{cf}$ is greatly influenced by the number of nodes. Given 16 validators in the committee, the latency could be 21.5 s, while the scenario with four nodes only introduces 1.4 s latency.

\subsection{Throughput Evaluation}

In the following set experiments, five Raspberry Pis work as validators in the committee to focus on throughput given limited influence from the committee size. To evaluate how much data could be processed during a certain period, the block confirmation time is calculated $\mathcal{T}_{bc}=(\mathcal{T}_{ct} + \mathcal{T}_{bp}+\mathcal{T}_{cf})$, which takes for microchain to complete an entire round of final-committee consensus with a varying block size between 512K and 4M. The block data throughput could be specified as $Th=\frac{\mathcal{T}_{bc}}{Block\_size} \times 3600$ (M/h), where M/h means Mbytes per hour. Test on varying block size, the result is: $Th_{512K}$=202 (M/h), $Th_{1M}$=293 (M/h), $Th_{2M}$=405 (M/h), and $Th_{4M}$=263 (M/h). Given fix transaction size like 1K, increasing the block size allows committing more transactions, and therefore reach a higher throughput, which maximizes the system capability. In the test, running microchain with 2M block size implies a theoretical maximum rate of  $\frac{405 \times 10^3K}{3600 \times 1K} \approx 113$ (tx/s). As block size increases, however, microchain achieves higher throughput at the cost of increased latency, and the throughput is constrained by network and system capability. For comparison, Bitcoin achieves a throughput of processing about seven transactions per second by committing a 2MB block per ten minutes.

\section{Conclusions}
\label{sec:conclusions}
Consensus is the core function of a blockchain system. This chapter introduces the basics of distributed consensus and identifies consensus goals in distributed systems. Given a comprehensive overview on blockchain consensus protocols in terms of BFT-based consensus, Nakamoto consensus and their varieties, challenges on integrating blockchain with IoT are evaluated. Finally, microchain is introduced as a case study that demonstrates the rationale and approach for designing scalable, lightweight blockchain protocols for IoT systems. 

The microchain provides a promising distributed ledger solution to IoT application scenarios. However, there remains a number of open issues in designing blockchain for IoT in term of security, scalability and efficiency. Although committee selection could improve the scalability of microchain, more investigation and test are needed to evaluate how committee selection algorithm scale to the network size. Another challenge is redesigning chain structure to address the ever-growing chain data size, which has a significant impact on computation and storage capability of IoT devices.

\bibliographystyle{spmpsci}
\bibliography{chapter10/references}

\begin{thebibliography}{10}
\providecommand{\url}[1]{{#1}}
\providecommand{\urlprefix}{URL }
\expandafter\ifx\csname urlstyle\endcsname\relax
  \providecommand{\doi}[1]{DOI~\discretionary{}{}{}#1}\else
  \providecommand{\doi}{DOI~\discretionary{}{}{}\begingroup
  \urlstyle{rm}\Url}\fi

\bibitem{ehtereum}
{Ethereum Homestead Documentation}.
\newblock \url{http://www.ethdocs.org/en/latest/index.html}

\bibitem{flask}
{Flask: A Pyhon Microframework}.
\newblock \url{http://flask.pocoo.org/}

\bibitem{litecoin}
{Litecoin - Open source P2P digital currency}.
\newblock \url{https://litecoin.org/}

\bibitem{pyca}
{pyca/cryptography documentation}.
\newblock \url{http://pyca/cryptography}

\bibitem{sqlite}
{SQLite}.
\newblock \url{https://www.sqlite.org/index.html}

\bibitem{blasch2019blockchain}
Blasch, E., Xu, R., Chen, Y., Chen, G., Shen, D.: Blockchain methods for
  trusted avionics systems.
\newblock arXiv preprint arXiv:1910.10638  (2019)

\bibitem{bonneau2015sok}
Bonneau, J., Miller, A., Clark, J., Narayanan, A., Kroll, J.A., Felten, E.W.:
  Sok: Research perspectives and challenges for bitcoin and cryptocurrencies.
\newblock In: 2015 IEEE Symposium on Security and Privacy, pp. 104--121. IEEE
  (2015)

\bibitem{buterin2017casper}
Buterin, V., Griffith, V.: Casper the friendly finality gadget.
\newblock arXiv preprint arXiv:1710.09437  (2017)

\bibitem{buterin2014next}
Buterin, V., et~al.: A next-generation smart contract and decentralized
  application platform.
\newblock white paper \textbf{3}, 37 (2014)

\bibitem{castro2002practical}
Castro, M., Liskov, B.: Practical byzantine fault tolerance and proactive
  recovery.
\newblock ACM Transactions on Computer Systems (TOCS) \textbf{20}(4), 398--461
  (2002)

\bibitem{castro1999practical}
Castro, M., Liskov, B., et~al.: Practical byzantine fault tolerance.
\newblock In: OSDI, vol.~99, pp. 173--186 (1999)

\bibitem{chen2018smart}
Chen, N., Chen, Y.: Smart city surveillance at the network edge in the era of
  iot: opportunities and challenges.
\newblock In: Smart Cities, pp. 153--176. Springer (2018)

\bibitem{croman2016scaling}
Croman, K., Decker, C., Eyal, I., Gencer, A.E., Juels, A., Kosba, A., Miller,
  A., Saxena, P., Shi, E., Sirer, E.G., et~al.: On scaling decentralized
  blockchains.
\newblock In: International Conference on Financial Cryptography and Data
  Security, pp. 106--125. Springer (2016)

\bibitem{eyal2016bitcoin}
Eyal, I., Gencer, A.E., Sirer, E.G., Van~Renesse, R.: Bitcoin-ng: A scalable
  blockchain protocol.
\newblock In: 13th $\{$USENIX$\}$ Symposium on Networked Systems Design and
  Implementation ($\{$NSDI$\}$ 16), pp. 45--59 (2016)

\bibitem{fischer1982impossibility}
Fischer, M.J., Lynch, N.A., Paterson, M.S.: Impossibility of distributed
  consensus with one faulty process.
\newblock Tech. rep., Massachusetts Inst of Tech Cambridge lab for Computer
  Science (1982)

\bibitem{gilad2017algorand}
Gilad, Y., Hemo, R., Micali, S., Vlachos, G., Zeldovich, N.: Algorand: Scaling
  byzantine agreements for cryptocurrencies.
\newblock In: Proceedings of the 26th Symposium on Operating Systems
  Principles, pp. 51--68. ACM (2017)

\bibitem{hopwood2016zcash}
Hopwood, D., Bowe, S., Hornby, T., Wilcox, N.: Zcash protocol specification.
\newblock Tech. rep. 2016--1.10. Zerocoin Electric Coin Company, Tech. Rep.
  (2016)

\bibitem{kiayias2017trees}
Kiayias, A., Panagiotakos, G.: On trees, chains and fast transactions in the
  blockchain.
\newblock In: International Conference on Cryptology and Information Security
  in Latin America, pp. 327--351. Springer (2017)

\bibitem{king2012ppcoin}
King, S., Nadal, S.: Ppcoin: Peer-to-peer crypto-currency with proof-of-stake.
\newblock self-published paper, August \textbf{19} (2012)

\bibitem{kokoris2018omniledger}
Kokoris-Kogias, E., Jovanovic, P., Gasser, L., Gailly, N., Syta, E., Ford, B.:
  Omniledger: A secure, scale-out, decentralized ledger via sharding.
\newblock In: 2018 IEEE Symposium on Security and Privacy (SP), pp. 583--598.
  IEEE (2018)

\bibitem{kwon2014tendermint}
Kwon, J.: Tendermint: Consensus without mining.
\newblock Draft v. 0.6, fall \textbf{1}, 11 (2014)

\bibitem{lamport1982byzantine}
Lamport, L., Shostak, R., Pease, M.: The byzantine generals problem.
\newblock ACM Transactions on Programming Languages and Systems (TOPLAS)
  \textbf{4}(3), 382--401 (1982)

\bibitem{lin2019enhance}
Lin, X., Xu, R., Chen, Y., Lum, J.: Enhance generalized exchange economy using
  blockchain: a time banking case study.
\newblock the IEEE Blockchain Technical Briefs  (2019)

\bibitem{lin2019blockchain}
Lin, X., Xu, R., Chen, Y., Lum, J.K.: A blockchain-enabled decentralized time
  banking for a new social value system.
\newblock In: 2019 IEEE Conference on Communications and Network Security
  (CNS), pp. 1--5. IEEE (2019)

\bibitem{liskov2012viewstamped}
Liskov, B., Cowling, J.: Viewstamped replication revisited  (2012)

\bibitem{liu2019survey}
Liu, Z., Luong, N.C., Wang, W., Niyato, D., Wang, P., Liang, Y.C., Kim, D.I.: A
  survey on applications of game theory in blockchain.
\newblock arXiv preprint arXiv:1902.10865  (2019)

\bibitem{luu2015scp}
Luu, L., Narayanan, V., Baweja, K., Zheng, C., Gilbert, S., Saxena, P.: Scp: A
  computationally-scalable byzantine consensus protocol for blockchains.
\newblock See https://www. weusecoins. com/assets/pdf/library/SCP  (2015)

\bibitem{luu2016secure}
Luu, L., Narayanan, V., Zheng, C., Baweja, K., Gilbert, S., Saxena, P.: A
  secure sharding protocol for open blockchains.
\newblock In: Proceedings of the 2016 ACM SIGSAC Conference on Computer and
  Communications Security, pp. 17--30. ACM (2016)

\bibitem{merkle1987digital}
Merkle, R.C.: A digital signature based on a conventional encryption function.
\newblock In: Conference on the theory and application of cryptographic
  techniques, pp. 369--378. Springer (1987)

\bibitem{nagothu2018microservice}
Nagothu, D., Xu, R., Nikouei, S.Y., Chen, Y.: A microservice-enabled
  architecture for smart surveillance using blockchain technology.
\newblock In: 2018 IEEE International Smart Cities Conference (ISC2), pp. 1--4.
  IEEE (2018)

\bibitem{nakamoto2008bitcoin}
Nakamoto, S., et~al.: Bitcoin: A peer-to-peer electronic cash system  (2008)

\bibitem{nikouei2019decentralized}
Nikouei, S.Y., Xu, R., Chen, Y., Aved, A., Blasch, E.: Decentralized smart
  surveillance through microservices platform.
\newblock In: Sensors and Systems for Space Applications XII, vol. 11017, p.
  110170K. International Society for Optics and Photonics (2019)

\bibitem{nikouei2018real}
Nikouei, S.Y., Xu, R., Nagothu, D., Chen, Y., Aved, A., Blasch, E.: Real-time
  index authentication for event-oriented surveillance video query using
  blockchain.
\newblock In: 2018 IEEE International Smart Cities Conference (ISC2), pp. 1--8.
  IEEE (2018)

\bibitem{novo2018blockchain}
Novo, O.: Blockchain meets iot: An architecture for scalable access management
  in iot.
\newblock IEEE Internet of Things Journal \textbf{5}(2), 1184--1195 (2018)

\bibitem{oki1988viewstamped}
Oki, B.M., Liskov, B.H.: Viewstamped replication: A new primary copy method to
  support highly-available distributed systems.
\newblock In: Proceedings of the seventh annual ACM Symposium on Principles of
  distributed computing, pp. 8--17. ACM (1988)

\bibitem{pease1980reaching}
Pease, M., Shostak, R., Lamport, L.: Reaching agreement in the presence of
  faults.
\newblock Journal of the ACM (JACM) \textbf{27}(2), 228--234 (1980)

\bibitem{ramachandran2018towards}
Ramachandran, G.S., Radhakrishnan, R., Krishnamachari, B.: Towards a
  decentralized data marketplace for smart cities.
\newblock In: 2018 IEEE International Smart Cities Conference (ISC2), pp. 1--8.
  IEEE (2018)

\bibitem{schneider1990implementing}
Schneider, F.B.: Implementing fault-tolerant services using the state machine
  approach: A tutorial.
\newblock ACM Computing Surveys (CSUR) \textbf{22}(4), 299--319 (1990)

\bibitem{schoenmakers1999simple}
Schoenmakers, B.: A simple publicly verifiable secret sharing scheme and its
  application to electronic voting.
\newblock In: Annual International Cryptology Conference, pp. 148--164.
  Springer (1999)

\bibitem{schwartz2014ripple}
Schwartz, D., Youngs, N., Britto, A., et~al.: The ripple protocol consensus
  algorithm.
\newblock Ripple Labs Inc White Paper \textbf{5}, 8 (2014)

\bibitem{shoker2017sustainable}
Shoker, A.: Sustainable blockchain through proof of exercise.
\newblock In: 2017 IEEE 16th International Symposium on Network Computing and
  Applications (NCA), pp. 1--9. IEEE (2017)

\bibitem{sompolinsky2016spectre}
Sompolinsky, Y., Lewenberg, Y., Zohar, A.: Spectre: A fast and scalable
  cryptocurrency protocol.
\newblock IACR Cryptology ePrint Archive \textbf{2016}, 1159 (2016)

\bibitem{stadler1996publicly}
Stadler, M.: Publicly verifiable secret sharing.
\newblock In: International Conference on the Theory and Applications of
  Cryptographic Techniques, pp. 190--199. Springer (1996)

\bibitem{syta2017scalable}
Syta, E., Jovanovic, P., Kogias, E.K., Gailly, N., Gasser, L., Khoffi, I.,
  Fischer, M.J., Ford, B.: Scalable bias-resistant distributed randomness.
\newblock In: 2017 IEEE Symposium on Security and Privacy (SP), pp. 444--460.
  Ieee (2017)

\bibitem{szabo1997formalizing}
Szabo, N.: Formalizing and securing relationships on public networks.
\newblock First Monday \textbf{2}(9) (1997)

\bibitem{vukolic2015quest}
Vukoli{\'c}, M.: The quest for scalable blockchain fabric: Proof-of-work vs.
  bft replication.
\newblock In: International workshop on open problems in network security, pp.
  112--125. Springer (2015)

\bibitem{wang2019survey}
Wang, W., Hoang, D.T., Hu, P., Xiong, Z., Niyato, D., Wang, P., Wen, Y., Kim,
  D.I.: A survey on consensus mechanisms and mining strategy management in
  blockchain networks.
\newblock IEEE Access \textbf{7}, 22328--22370 (2019)

\bibitem{xiao2019distributed}
Xiao, Y., Zhang, N., Li, J., Lou, W., Hou, Y.T.: Distributed consensus
  protocols and algorithms.
\newblock Blockchain for Distributed Systems Security p.~25 (2019)

\bibitem{xiao2019survey}
Xiao, Y., Zhang, N., Lou, W., Hou, Y.T.: A survey of distributed consensus
  protocols for blockchain networks.
\newblock arXiv preprint arXiv:1904.04098  (2019)

\bibitem{xu2019decentralized}
Xu, R., Chen, S., Yang, L., Chen, Y., Chen, G.: Decentralized autonomous
  imaging data processing using blockchain.
\newblock In: Multimodal Biomedical Imaging XIV, vol. 10871, p. 108710U.
  International Society for Optics and Photonics (2019)

\bibitem{xu2018blendcac}
Xu, R., Chen, Y., Blasch, E., Chen, G.: Blendcac: A blockchain-enabled
  decentralized capability-based access control for iots.
\newblock In: 2018 IEEE International Conference on Internet of Things
  (iThings) and IEEE Green Computing and Communications (GreenCom) and IEEE
  Cyber, Physical and Social Computing (CPSCom) and IEEE Smart Data
  (SmartData), pp. 1027--1034. IEEE (2018)

\bibitem{xu2018smartcac}
Xu, R., Chen, Y., Blasch, E., Chen, G.: Blendcac: A smart contract enabled
  decentralized capability-based access control mechanism for the iot.
\newblock Computers \textbf{7}(3), 39 (2018)

\bibitem{xu2018exploration}
Xu, R., Chen, Y., Blasch, E., Chen, G.: Exploration of blockchain-enabled
  decentralized capability-based access control strategy for space situation
  awareness.
\newblock Optical Engineering \textbf{58}, 58 -- 58 -- 16 (2019).
\newblock \doi{10.1117/1.OE.58.4.041609}.
\newblock \urlprefix\url{https://doi.org/10.1117/1.OE.58.4.041609}

\bibitem{xu2019microchain}
Xu, R., Chen, Y., Blasch, E., Chen, G.: Microchain: A hybrid consensus
  mechanism for lightweight distributed ledger for iot.
\newblock arXiv preprint arXiv:1909.10948  (2019)

\bibitem{xu2018constructing}
Xu, R., Lin, X., Dong, Q., Chen, Y.: Constructing trustworthy and safe
  communities on a blockchain-enabled social credits system.
\newblock In: Proceedings of the 15th EAI International Conference on Mobile
  and Ubiquitous Systems: Computing, Networking and Services, pp. 449--453. ACM
  (2018)

\bibitem{xu2019blendmas}
Xu, R., Nikouei, S.Y., Chen, Y., Blasch, E., Aved, A.: Blendmas: A
  blockchain-enabled decentralized microservices architecture for smart public
  safety.
\newblock In: The 2019 IEEE International Conference on Blockchain
  (Blockchain-2019), pp. 1--8. IEEE (2019)

\bibitem{xu2019blendsm}
Xu, R., Ramachandran, G.S., Chen, Y., Krishnamachari, B.: Blendsm-ddm:
  Blockchain-enabled secure microservices for decentralized data marketplaces.
\newblock In: 2019 IEEE International Smart Cities Conference (ISC2). IEEE
  (2019)

\bibitem{zamani2018rapidchain}
Zamani, M., Movahedi, M., Raykova, M.: Rapidchain: Scaling blockchain via full
  sharding.
\newblock In: Proceedings of the 2018 ACM SIGSAC Conference on Computer and
  Communications Security, pp. 931--948. ACM (2018)

\bibitem{zhang2017rem}
Zhang, F., Eyal, I., Escriva, R., Juels, A., Van~Renesse, R.: $\{$REM$\}$:
  Resource-efficient mining for blockchains.
\newblock In: 26th $\{$USENIX$\}$ Security Symposium ($\{$USENIX$\}$ Security
  17), pp. 1427--1444 (2017)

\end{thebibliography}

\end{document}